\documentclass[fleqn,usenatbib]{mnras}

\usepackage[T1]{fontenc}
\usepackage{ae,aecompl}
\usepackage{amsmath}
\usepackage{float}
\usepackage{booktabs}
\usepackage{ulem}

\usepackage{graphicx}	
\usepackage{amsmath}	
\usepackage{amssymb}	
\usepackage{float}		
\usepackage{multicol}   

\begin{document}
\title[Asteroseismology of {\it SZ Lyn} using Multi-Band High Time Resolution Photometry ]{Asteroseismology of {\it SZ Lyn} using Multi-Band High Time Resolution Photometry from Ground and Space}

\author[J. Adassuriya et al.]{J. Adassuriya$^{1}$\thanks{E-mail:adassuriya@gmail.com}, S. Ganesh$^{2}$, J. L. Guti\'errez$^{3}$, G. Handler$^{4}$, Santosh Joshi$^{5}$
\newauthor{K. P. S. C. Jayaratne$^{6}$, K. S. Baliyan$^{2}$}\\
$^{1}$Astronomy Division, Arthur C Clarke Institute, Katubedda, Moratuwa, Sri Lanka\\
$^{2}$Astronomy and Astrophysics Division, Physical Research Laboratory, Ahmedabad, India\\
$^{3}$Department of Physics, Universitat Polit\`ecnica de Catalunya, Castelldefels, Spain \\ 
$^{4}$Nicolaus Copernicus Astronomical Center, Bartycka 18, 00-716 Warsaw, Poland\\
$^{5}$Aryabhatta Research Institute of Observational Sciences (ARIES), Manora Peak-Nainital, India\\
$^{6}$Department of Physics, University of Colombo, Colombo, Sri Lanka}

\date{Accepted XXX. Received YYY; in original form ZZZ}

\pubyear{2020}

\label{firstpage}
\pagerange{\pageref{firstpage}--\pageref{lastpage}}

\maketitle

\begin{abstract}
We report the analysis of high temporal resolution ground and space based photometric observations of SZ Lyncis, a binary star one of whose components is a high amplitude $\delta$ Scuti. UBVR photometric  observations were obtained from Mt. Abu Infrared Observatory and Fairborn Observatory; archival observations from the WASP project were also included. Furthermore, the continuous, high quality light curve from the TESS project was extensively used for the analysis. The well resolved light curve from TESS reveals the presence of 23 frequencies with four independent modes, 13 harmonics of the main pulsation frequency of $8.296943\pm0.000002$ d$^{-1}$ and their combinations. The frequency 8.296 d$^{-1}$ is identified as the fundamental radial mode by amplitude ratio method and using the estimated pulsation constant. The frequencies 14.535 d$^{-1}$, 32.620 d$^{-1}$ and 4.584 d$^{-1}$ are newly discovered for SZ Lyn. Out of these three, 14.535 d$^{-1}$ and 32.620 d$^{-1}$ are identified as non-radial lower order \textit{p}-modes and 4.584 d$^{-1}$ could be an indication of a \textit{g}-mode in a $\delta$ Scuti star. As a result of frequency determination and mode identification, the physical parameters of SZ Lyn were revised by optimizations of stellar pulsation models with the observed frequencies. The theoretical models correspond to 7500 K $\le$T$_{\rm eff}$ $\le$ 7800 K, log(g)=3.81$\pm0.06$. The mass of SZ Lyn was estimated to be close to 1.7--2.0 M$_\odot$ using evolutionary sequences. The period-density relation estimates a mean density $\rho$ of 0.1054$\pm0.0016$ g cm$^{-3}$.   
\end{abstract}




\section{Introduction}

The study of multi-periodic stellar pulsations represents a unique methodology for studying the interior of a star \citep{2010aste.book.....A, aerts2019}. Stellar pulsations can be either radial and/or non-radial in nature and is driven through either pressure waves (\textit{p}-mode) or gravity waves (\textit{g}-mode). In \textit{p}-modes,  pressure is the primary restoring force for a  star perturbed from equilibrium, while in \textit{g}-mode, buoyancy is the restoring force \citep{2010aste.book.....A}. Diverse pulsation group includes Beta Cephei ($\beta$ Cep), slowly pulsating B (SPB) stars, gamma Doradus ($\gamma$ Dor), $\delta$ Scuti stars etc. \citet{Garrido1990, 2000ASPCBreger, Handler2006, Handler2008} provide detailed discussion on the ground based observations of these stars. The group of $\delta$ Scuti stars is located in the region where the classical instability strip intersects to the main sequence in the Hertzsprung-Russell (HR) diagram \citep[e.g.,][]{murphy2019}. Their spectral range is from A2 to F0 corresponding to temperatures of 7000 K < T$_{\rm eff}$ < 9300 K \citep{uytterhoeven2011}. Typical periods for \textit{p}-mode pulsations in $\delta$ Scuti stars range from 15 min to 5 h \citep{uytterhoeven2011}.
$\delta$ Scuti stars which simultaneously pulsate in \textit{p}-modes and gravity \textit{g}-modes excited by the $\kappa$ mechanism and the convective flux blocking mechanism, respectively, are commonly referred to as hybrid stars \citep{bowman2018mnras}.
These stars are of intermediate mass ranging between 1.5 to 2.5 M$_{\odot}$ \citep{2010aste.book.....A}, hence most of them are in the core hydrogen burning or shell hydrogen burning phase and possess convective cores. (See \citet{2000ASPCBreger} for a detailed discussion on these stars). Furthermore, $\delta$ Scuti type pulsations have been detected in many pre- and post-main sequence \citep{antoci2019} stars while low amplitude $\delta$ Scuti type pulsations have been detected in many metallic A-type (Am) stars \citep[e.g.,][]{2012MNRAS.424.2002J,2016A&A...590A.116J, 2017MNRAS.467..633J, 2017MNRAS.465.2662S}. The complex interior structures of these stars can only be examined by a study of the multi-periodic radial and non-radial pulsation modes, a technique known as asteroseismology \citep{2010aste.book.....A,2015JApA...36...33J}. Generally, the frequency spectrum of a $\delta$ Scuti star is consisting of both \textit{g} and \textit{p} modes with many harmonics and combinations. \citet{breger2011}, \citet{murphy2013}, \citet{papics2012gravito} provide comprehensive analysis on frequency combinations and harmonics of $\delta$ Scuti targets. 

In this paper, we present a detailed analysis of SZ Lyncis, HD 67390 (RA=08h 09m 35.8s, DEC=+44$^\circ{}$ 28${'}$ 17.6${''}$) a High Amplitude $\delta{}$ Scuti (HADS) type binary star, m$_{v}$= 9.1 mag, with reported pulsation and orbital period 0.12053793 days and 1173.5 days, respectively \citep{Soliman1986}. SZ Lyn is the brighter component of a binary system where the fainter component is not observed with spectroscopic techniques, and is hence characterized as a single line spectroscopic binary \citep{Gazeas2004}. The fundamental pulsation period of this star was first determined by \citet{Binnendijk1968} and later refined by \citet{Gazeas2004}.

SZ Lyn has been studied on a number of occasions for the orbital and pulsation parameters. \citet{Van1967} reported that the linear ephemeris for the time of pulsation maximum has not been accurately estimated. \citet{Barnes1975} suggested that the very long period orbital motion of SZ Lyn affected the linear ephemeris due to the light-travel time across the orbit. Using photometric and spectroscopic data, \citet{Moffett1988orbital} computed non-linear ephemeris for the times of light maxima and determined improved values for the pulsation and orbital parameters. A similar analysis was made by \citet{Paparo1988} and \citet{Qian2013}. The observed minus calculated (O-C) diagrams show periodic as well as secular variations, and \citet{Paparo1988} concluded that the main pulsation period changes by $(2.25\pm0.42)\times 10^{-12}$ day per cycle. This value was later refined to $(2.90\pm0.22)\times 10^{-12}$ day per cycle by \citet{Gazeas2004}.        

\begin{table}
    \centering
    \caption{Physical parameters of SZ Lyn derived in  literature. The parameters 
    were used to generate pulsation and evolutionary models of SZ Lyn.
    \label{tab:params}
    }
    \begin{tabular}{l|l|l}
    \hline
    Parameter & Value & Reference\\
    \hline
     T$_{\rm eff}$ (K)  & 7540  & \citet{Langford1976}\\
                & 7235  & LAMOST \\
                & 7799  & Gaia \\
                & 6830$\pm$150 & \citet{johnson1966} \\
    $\log(g)$   & 3.88 & \citet{Langford1976}\\
                & 3.94 & LAMOST \\
        Mass (M$_\odot$) & 1.57$^{+0.17}_{-0.66}$ & \citet{Fernley1983} \\
        Radius (R$_\odot$) & 2.76$^{+0.11}_{-0.46}$ & \citet{Fernley1983} \\
                & 2.80 & \citet{1981AA} \\
        Parallax (mas) & 2.49$\pm$0.07 & Gaia \\
    \hline    
    \end{tabular}
\end{table}

The physical parameters of SZ Lyn determined by several authors are given in Table \ref{tab:params}. This star has nearly solar abundance
\citep{Alania1972, Langford1976}. 
\citet{1981AA} determined the radial velocity of SZ Lyn due to the pulsation as 30 km s$^{-1}$ using high time resolution radial velocity curves. 
\textit{Gaia} \citep{2016A&A, 2018A&A, 2018A&A9Luri} determined a parallax of $2.49\pm0.07$ milli-arc seconds, which, according to \citet{bailer2018} places it at a distance of $397\pm11$ pc. Most of the previous studies of SZ Lyn reported the main pulsation period and a few attempts have also been made to search for additional harmonics and secondary pulsation periods. Apart from the main pulsation period, two harmonics of the fundamental period were reported by \citet{Gazeas2004}.  

We have subjected SZ Lyn to asteroseismic analysis using high temporal resolution, multi-band light curves obtained from the ground, as well as highly-precise photometric data from the TESS space mission; we attempted to find more frequencies in SZ Lyn and identify the pulsation modes, basically \textit{n} and \textit{l}. The manuscript is organised as follows: The observational data and reduction procedures are given in Sec. \ref{obs}. The frequency analysis is performed in Sec. \ref{freq}. In Sec. \ref{mode}, the mode identification procedures are discussed in detail paying special attention to the theoretical model analysis. The refinement of stellar parameters of SZ Lyn is presented in Sec. \ref{para}, followed by the evolutionary and pulsation models. Finally, we have discussed the results with a detailed consideration of the limitations and uncertainties of our analysis in Sec. \ref{discuss} and summarized the results under Sec. \ref{conclusion}.

\section{Photometric Observations}
\label{obs}
Photometry data from various sources (Mount Abu, APT, WASP and TESS) are included in the current study. The Table \ref{tab:obslog} compiles the observing log along with details of the filters used, the cadence, number of nights, and the total number of data points.  In the following subsections, the data are further described. The observed ground based data are inconsistent with respect to the comparison stars and magnitude scales. Due to these reasons, we refrained from combining all three data sources into a single light curve and hence the analyses were performed separately.

\subsection{Mount Abu}

A series of observations were carried out from Mt. Abu InfraRed Observatory (MIRO). The observatory is located at the highest peak, Gurushikhar, of Aravali range near Mount Abu in the western state of Rajasthan, India. 
The observatory is located at 24.65$^\circ{}$ N, 72.78$^\circ{}$ E, at an altitude of 1680 meters where one can obtain typical seeing of $\sim $1.2 arc seconds.

Observations were made with the Corrected Dall-Kirkham (CDK) 50 cm, f/6.8 equatorial mount telescope equipped with an Andor 1024$\times{}$1024 EMCCD, thermo-electrically cooled to $-80 ^\circ{}$C. Details about the setup are provided by \citet{Ganesh2013}. 
The higher frame rates and negligible readout noise ($<$1 electron with EM gain), characteristic of EMCCD, are ideal for the observation of short period variable stars with high cadence. The large EMCCD array provides a field of view of 13$\times{}$13 arc minutes.
The system was fully automated and the exposures were scripted using RTS2\footnote{\label{rts2}RTS2 : Remote Telescope System v2 is available at \url{http://rts2.org/}} which allowed to obtain repeated sequences of B, V and R single frames with different exposure times for each filter. The filter movement being automated, very little time is lost in this process. After removing faulty frames, we have acquired a total of 4328 frames in B, 4569 in V and 6836 in R during the seven nights of observation between December 2013 and November 2016 (see Table \ref{tab:obslog}). At least one pulsation cycle of SZ Lyn was covered in B, V and R on each night.\\ 

\begin{table*}
\begin{center}
\caption{Photometric observation of SZ Lyn. The JD column shows exact date of observation at Mount Abu and the observation periods of the other three data sources.}
\label{tab:obslog}
\begin{tabular}{ c | c | c | r | r | r | r }
 \hline
 \hline
Observation & JD (2450000) & Effective nights & Band & Coverage (h) & Cadence (s) & Data points \\
 \hline
 Mount Abu & 6639.104 & 7 & B & 29.2 & 5 & 4328 \\
 (India) & 6640.153 & & V & 28.9 & 2 & 4569 \\
 & 6664.134 & & R & 55.3 & 2 & 6836 \\
 & 6684.118 & & & & & \\
 & 6686.174 & & & & & \\
 & 6693.097 & & & & & \\
 & 7705.135 & & & & & \\
 \hline
 APT & 7480.634-7515.688 & 24 & U & 50.1 & 60 & 701 \\
 (USA) & & & B & 51.4 & 40 & 688 \\
 & & & V & 52.5 & 40 & 672 \\
 \hline
 WASP & 4190.366-4575.408 & 39 & (400-750) nm & 216.0 & 60 & 2894 \\
 \hline
 TESS & 8842.503-8868.827 & 25 (days) & (600-1000) nm & 631.8 & 120 & 16550 \\
 \hline
\end{tabular}
\end{center}
\end{table*}

The basic reduction procedures of the Mount Abu data were performed using the \textsc{IRAF}\footnote{\textsc{IRAF} is distributed by the National Optical Astronomy Observatory, which is operated by the Association of Universities for Research in Astronomy (AURA) under a cooperative agreement with the National Science Foundation.} software. The photometry was performed by running the \textit{phot} package in \textsc{IRAF}. The instrumental magnitudes were extracted for the BVR bands using an aperture of size 3 to 4 times the FWHM of program star for different filters. An annulus of 5 pixels width was used for all the bands to determine the background. Two stars of magnitudes m$_{V}$= 10.5 (TYC 2979-1329) and m$_{\rm V}$= 10.8 (TYC 2979-1343), in the same field of view were used as comparison stars for differential photometry in each band.\

The colour variations (B-V) and (V-R) were determined for the Mount Abu data. The amplitude of the variations of SZ Lyn in the BVR bands were $\Delta$B$\approx$0.7, $\Delta$V$\approx$0.5 and $\Delta$R$\approx$0.4 magnitudes. Both colour and magnitude variations reveal that magnitude change in B is the highest, as expected for a $\delta$ Scuti star. Therefore, B band was used as the reference filter in the UBVR system for the calculation of observed and theoretical amplitude ratios presented in Sec. \ref{mode}.\\

\begin{figure*}
\centering
\includegraphics[width=514pt, height=258pt]{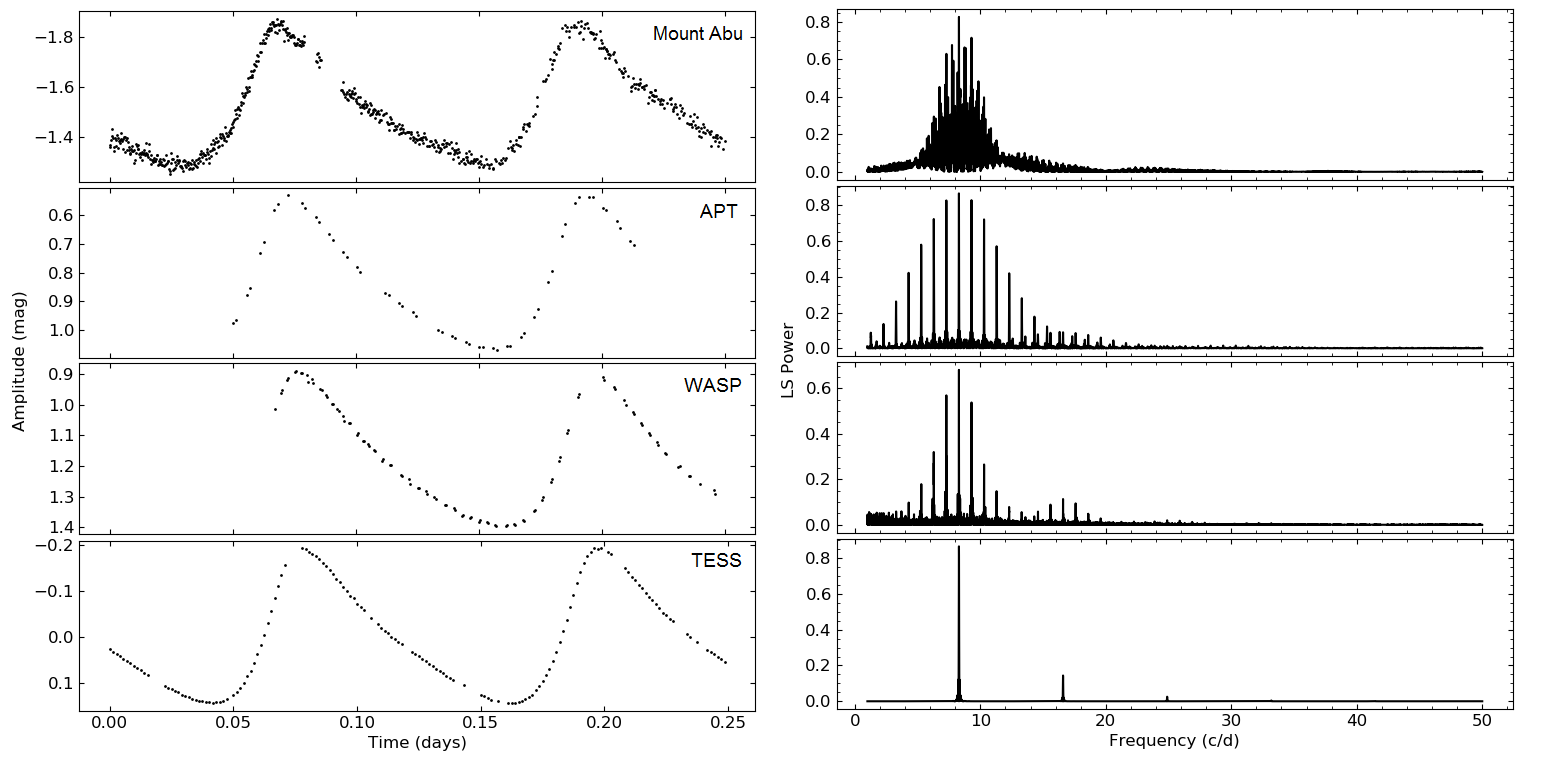}
\caption{The high cadence light curves of ground and space observations for 0.25 day coverage of SZ Lyn. The phases are synchronized at the peaks and hence the WASP and APT data do not start at 0. The power spectra in the frequency range 2 - 50 d$^{-1}$ is shown in the right panel corresponding to each data source. V band light curves are shown for the Mount Abu and APT data. The amplitude of the light curves are in magnitude scale where the ground based data is differential and TESS fluxes are converted to magnitudes.}
\label{fig:BVR_magnitudes.jpg_figure}
\end{figure*}

\subsection{APT}

Differential time-series photoelectric data were collected
through the Johnson UBV filters with the 0.75 m Automatic Photoelectric
Telescope (APT) T6 at Fairborn Observatory in Arizona \citep{1997PASP}. 
Two comparison stars, HD 67808 and HD 66113,
were used. These stars are too distant from SZ Lyn that they would have fitted onto the Mt. Abu CCD frames, but were chosen to have colours and magnitudes similar to those of the target. Integration times were 2 $\times$ 20 s in each filter, with the
exception of the U filter for which an integration time of 2 $\times$ 30 s was used to collect enough photons.

The data were reduced following standard photoelectric photometry
schemes. First, the measurements were corrected for coincidence losses.
Then, sky background was subtracted within each target/local comparison
star group. Standard extinction coefficients were employed; small errors
in their assumption would be compensated when computing differential
magnitudes. Obviously, the same extinction correction was applied to
each star. Some bad measurements due to partly poor telescope tracking
had to be eliminated. Finally, differential magnitudes were computed by
interpolation, and the timings were converted to Heliocentric Julian
Date. During the reductions it turned out that HD 66113 could be a
low-amplitude $\delta$ Scuti star ($f=12.573$ d$^{-1}$, $A_V=2.1$ mmag), hence
it was rejected from the computation of the differential light curves of
SZ Lyn. A total of 701/688/672 good
measurements for the U, B, and V filters, respectively, were obtained with an
estimated accuracy of 4.5/3.5/3.5 mmag per point. The time span of the
combined data set is 35.05 d, with data collected on 24 nights.\

\subsection{WASP}  

The Wide Angle Search for Planets (WASP) observatories consist of two identical robotic telescopes, one located at La Palma and other at South African Astronomical Observatory \citep{Pollacco2006}. The WASP program also observed SZ Lyn during the period March 2007 to April 2008 in its own wide pass-band (Table \ref{tab:obslog}). During routine observations, two exposures of each field with 30 seconds were obtained, and each field was sampled every 9--12 minutes; for SZ Lyn that resulted in a total of 2894 data points equivalent to 216 effective hours of observation. The flux were converted to magnitudes by the WASP pipeline \citep{2010Butters} so that the light curve in Fig. \ref{fig:BVR_magnitudes.jpg_figure} is in magnitude scale. The WASP data are neither equally distributed nor differential, and the differential magnitudes from the APT data were computed with respect to a much brighter and bluer star than Mount Abu data.\ 

\subsection{TESS}
SZ Lyn was observed by TESS \citep{ricker2015} in sector 20, ID number TIC 192939152, from 24$^{\rm th}$ December 2019 to 19$^{\rm th}$ January 2020 in 120 seconds of cadence. The light curves were generated using simple aperture photometry and corrected by pre-search data conditioning (PDC). The PDC pipeline module uses singular value decomposition to identify and correct for time-correlated instrumental signatures such as, spacecraft pointing jitter, long-term pointing drifts due to differential velocity aberration, and other stochastic errors \citep{jenkins2016,balona2019}. The corrected flux from TESS Asteroseismic Science Operation Center (TASOC) were converted to magnitudes. A part of light curve of each data source for the duration of 0.25 days is shown in Fig. \ref{fig:BVR_magnitudes.jpg_figure}. The power spectra on the right side in the same figure are a good illustration of how the nearly uninterrupted space data result in a much cleaner spectral window function compared to the ground based data. 
\section{Frequency Analysis}
\label{freq}
For seismic analysis, the light curves observed from  Mount Abu, APT WASP and TESS data were searched for frequencies independently. We used the generalized Lomb-Scargle (LS) \citep{2009A&A} algorithm since ground based observations were highly unevenly sampled and LS is specially designed to pick out periodic variations in such kind of data. The LS periodogram tool is available in the \textsc{VARTOOLS} software package \citep{hartman2012vartools}. 
The highest frequency to search is the Nyquist frequency (\textit{f$_{\rm Nq}$}) and should be derived from the data set. For TESS data this is approximately 360 d$^{-1}$. \textit{f$_{\rm Nq}$} of the ground based observations differ due to the uneven observation parameters. 
As we could not detect frequencies above 120 d$^{-1}$ up to the TESS upper limit of 360 d$^{-1}$, we limit our search to the range 0 - 120 d$^{-1}$ where frequency peaks satisfying a SNR>4 significance criterion \citep{Breger1993} occur in both space and ground based data. In LS analysis, the frequency step size is depending on 1/T, where T is the time span of the observation. The parameter 1/T in frequency analysis is termed as Rayleigh criterion \citep{2010aste.book.....A}. When T is higher the peaks of the power spectrum are narrower. Though the ground based observation spreads over long range of time, the desired narrowness could not be achieved due to the low duty cycle. This drastic difference in ground and space based power spectra is clearly seen in Fig. \ref{fig:BVR_magnitudes.jpg_figure}. Due to the different T in the four data sources, the sampling frequency of LS analysis are different but we ensured that the sampling frequencies were kept more than the highest frequency we expected to find. The  light curves were whitened at each peak and applied 5$\sigma$ clipping in the calculation of the average and RMS of the power spectrum when computing the SNR value of a peak. The power in LS periodogram is normalized to unity \citep{2009A&A} so that the frequency peaks of the four data sources can be compared graphically. Due to the unavailability of phase information of frequencies in LS analysis, we performed the Discrete Fourier Transformation (DFT) using \textsc{Period04} \citep{2004IAUS} keeping the same searching parameters as in LS method and obtained amplitudes and phases of the frequencies. The power spectra for the frequency range of 0 - 120 d$^{-1}$ were obtained by both methods. The frequencies determined by both methods are consistent. \textsc{Period04} determined the error of the frequencies, amplitudes and phases using 100 iterations of Monte Carlo simulation.  Due to the single-site observation, the power spectra of the ground based observations were severely affected by $\pm1$ d$^{-1}$ aliasing and its multiples, $\pm2$ d$^{-1}$, $\pm3$ d$^{-1}$ etc. The power spectra in the range of 0 - 100 d$^{-1}$ for three data sources are shown in Fig. \ref{fig:powerspectra_ground.png_figure}. For clarity and save space we exclude the power spectra of WASP data in Fig. \ref{fig:powerspectra_ground.png_figure}.\

The identified frequencies of both space and ground based observations were tabulated in Table \ref{frequencies of SZ Lyn}. With the long time base observations of TESS, 23 frequencies are identified, while up to 10 of them are discovered in the Mount Abu BVR bands, 8 in the APT UBV bands and 5 in WASP data. Though the frequency extraction from the ground based observations is more complicated due to the low observation span and gaps, we used pre-whitening and the \textit{killharm} routine in \textsc{VARTOOLS} to carefully remove the fundamental frequency of \textit{f$_{1}$} 8.296 d$^{-1}$ and its first three harmonics (corresponding to frequencies of 16.590, 24.890 and 33.187 d$^{-1}$), as well as the $\pm1$ d$^{-1}$ day aliasing. 
The first four frequencies were removed from the Mount Abu data using their Fourier components and the residual light curve was obtained.  
The residuals were again subjected to LS and DFT. The rest of the multiples (5$\times$\textit{f$_{1}$} to 8$\times$\textit{f$_{1}$}) of ground based data were found using these residual light curves. Eight common frequencies were clearly identified in Mount Abu and APT power spectra while WASP detected only up to five frequencies. The main pulsation frequency \textit{f$_{1}$} and its seven harmonics were common in space and ground observations. In addition, there are six more harmonics of \textit{f$_{1}$} present in TESS data. The harmonics above \textit{f$_{3}$} are newly discovered for SZ Lyn in this work. The \textit{f$_{1}$} of 8.296 d$^{-1}$, that we identify as the fundamental radial mode, had previously been identified by \citet{Gazeas2004} and is further confirmed by the asteroseismic technique of amplitude ratio method in the Sec. \ref{mode}. \ 

\subsection{New frequencies of SZ Lyn}
\begin{figure*}
\includegraphics[width=450pt,height=260pt]{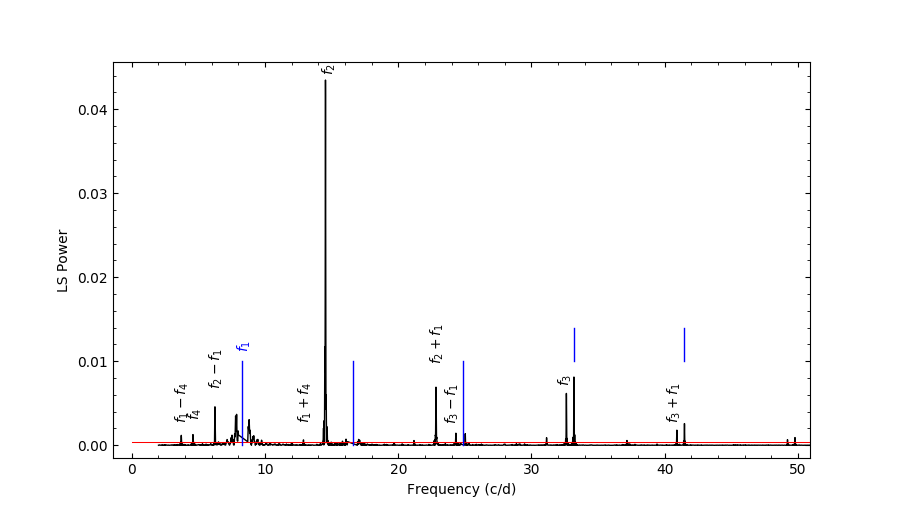}
\caption{New set of frequencies and their combinations with the fundamental frequency \textit{f$_{1}$} detected in SZ Lyn using TESS data. The blue lines indicate the location of the whitened frequencies, \textit{f$_{1}$} and its first four harmonics. The discontinuity at the frequency \textit{f$_{1}$} is due to the removal of strong side-lobes created by \textit{f$_{1}$}. The horizontal red line is the $4\sigma$ threshold line for cut off.}
\label{fig:powerspectra_space.png_figure}
\end{figure*}

Most importantly, the frequencies \textit{f$_{2}$} (14.535 d$^{-1}$), \textit{f$_{3}$} (32.620 d$^{-1}$) and \textit{f$_{4}$} (4.584 d$^{-1}$), none of them harmonics of \textit{f$_{1}$}, are new for SZ Lyn. The truncated frequency spectrum in Fig. \ref{fig:powerspectra_space.png_figure} shows the linear combinations present due to the frequencies \textit{f$_{1}$}, \textit{f$_{2}$}, \textit{f$_{3}$} and \textit{f$_{4}$}. These linear combinations in one way support that the newly discovered frequencies, \textit{f$_{2}$}, \textit{f$_{3}$} and \textit{f$_{4}$}, are independent modes. The possibility of these three frequencies being combinations of higher frequency higher order \textit{l} modes is also minimum due to lack of frequencies in higher region of the frequency spectrum of TESS. Therefore, we further analyze \textit{f$_{2}$}, \textit{f$_{3}$} and \textit{f$_{4}$} assuming they are base frequencies. Fig. \ref{fig:powerspectra_space.png_figure}  shows the combinations made by \textit{f$_{2}$}, \textit{f$_{3}$} with the well defined radial fundamental frequency \textit{f$_{1}$}. In fact, radial overtones are common in $\delta$ Scuti stars, and so we could in principle consider \textit{f$_{2}$} as an overtone of fundamental. Nevertheless, we estimated the first and second overtones for the well defined period ratios of [0.772 - 0.776] range of \citet{suarez2006} and [0.611 - 0.632] range of \citet{Stelling1979}, respectively. Given that the fundamental radial mode is 8.296 d$^{-1}$, the calculated first and second overtones are 10.746 and 13.578 d$^{-1}$ which are not found in the observed power spectra and \textit{f$_{2}$} is too far from these two overtones as well. Therefore, it is possible to conclude \textit{f$_{2}$} is a nonradial mode. To determine the spherical degree, \textit{l}, of the nonradial mode, rotational splitting is widely used \citep{kurtz2014, breger2011}. We noticed that there are incomplete frequency multiplets around \textit{f$_{2}$}, one at 14.503 and the other 14.435 d$^{-1}$. Assuming this can be a sign of rotational splitting and in order to confirm it, we fitted the 14.503 and 14.435 d$^{-1}$ and pre-whitened. The residuals still show several peaks which often occur when a signal changes its amplitude and frequency slightly during the course of the observations. Fourier decomposition of such a signal produces peaks that are real, but resemble rotational splitting \citep{2002A&A...385..537B,2016MNRAS.460.1970B}. Another point supporting in this direction is that the combination of \textit{f$_{2}$} with the main pulsation, \textit{f$_{1}$}, at 22.830 d$^{-1}$, also shows similar asymmetry. This is another clue, although not unambiguous, that we see the beating of a time-variable pulsation signal with the main radial mode.\

The frequency \textit{f$_{4}$} is located in the low frequency range and less than the assumed fundamental radial mode of \textit{f$_{1}$}. \citet{kurtz2014}, \citet{2010aste.book.....A}, \citet{breger199930+} strongly state that gravity modes (\textit{g}-modes) are located in the range 0 - 5 d$^{-1}$. Therefore, the frequency \textit{f$_{4}$} could be a medium order gravity mode. Based on arguments given  \citet{kurtz2015}, \citet{kurtz2014}, \textit{f$_{4}$} could be a combination of higher order \textit{p}-modes or very low order \textit{g}-modes.
Furthermore, \citet{kurtz2014} show that the availability of combinations $\nu_{1} \pm \nu_{g}$ with $\nu_{1}$ and $\nu_{g}$ being the frequencies of the highest amplitude singlet \textit{p}-mode and a \textit{g}-mode respectively. These combinations of frequencies produced by fundamental radial \textit{p}-mode and a \textit{g}-mode can be naturally explained by non-linear effects that occur when the \textit{p}-mode and the \textit{g}-modes are excited simultaneously. The combination frequencies of the suspected \textit{g}-mode with the dominant frequency  are given by \textit{f$_{1}+$f$_{4}$} (12.880 d$^{-1}$) and \textit{f$_{1}-$f$_{4}$} (3.703 d$^{-1}$) which are symmetrically placed around the fundamental \textit{p}-mode frequency (\textit{f$_{1}$}) in Fig. \ref{fig:powerspectra_space.png_figure}. The availability of these combinations in SZ Lyn supports the idea that \textit{f$_{4}$} is a gravity mode and indicates that both \textit{p}-modes and \textit{g}-modes are present in SZ Lyn.  

\begin{figure*}
\includegraphics[width=0.9\textwidth]{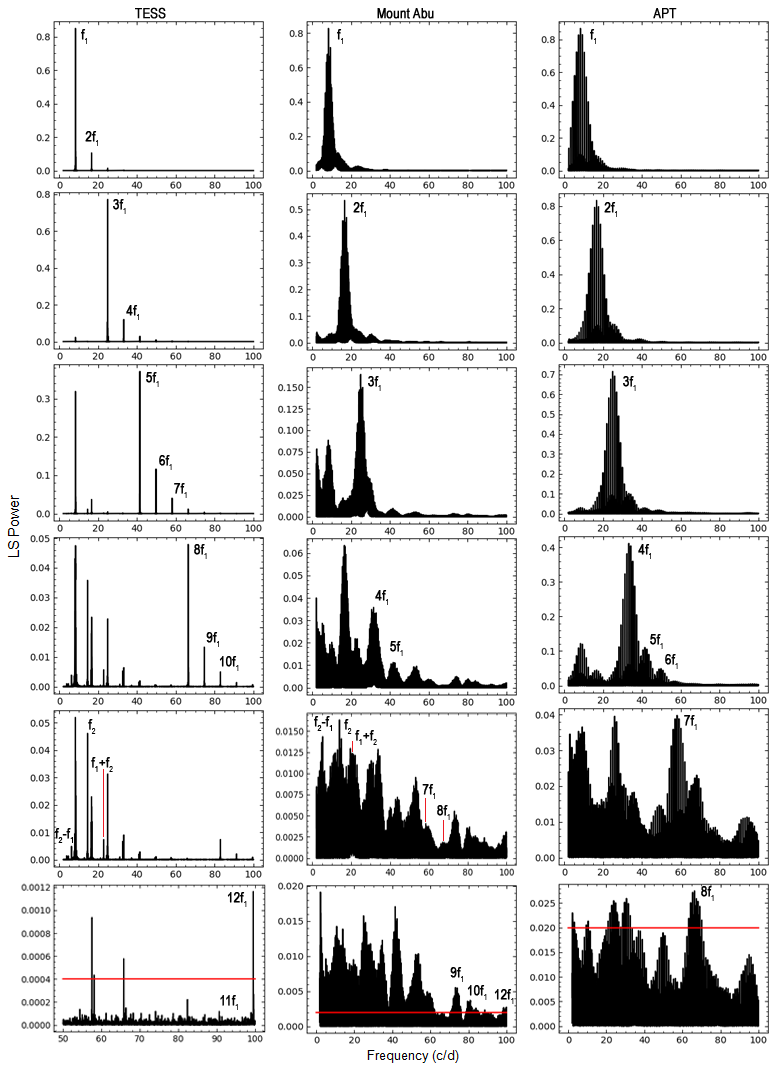}
\caption{The comparison of LS power spectra of space and ground base observation. The low amplitude frequencies were recovered by whitening the previously determined frequencies f$_1$, 2f$_1$, 3f$_1$ etc. The horizontal red lines in the last row indicates the $4\sigma$ threshold for three data sources.}
\label{fig:powerspectra_ground.png_figure}
\end{figure*}

Due to the lower precision of the ground-based data we could not clearly detect the three frequencies, \textit{f$_{2}$}, \textit{f$_{3}$} and \textit{f$_{4}$} in these data. However, the power spectra of Mount Abu were able to just resolve \textit{f$_{2}$} peak despite the crowded noisy field (See Fig. \ref{fig:powerspectra_ground.png_figure}). But the amplitudes of \textit{f$_{2}$} determined in UBVR colour bands of ground based data are highly unreliable; therefore, we could not include \textit{f$_{2}$} in Sec. \ref{mode} to determine the spherical degree \textit{l} using amplitude ratio method. The amplitudes $A$ and the phases $\phi$ of the frequency \textit{f$_1$} determined for UBVR colour bands using \textsc{Period04} are shown in Table \ref{observational amplitudes}.\

\begin{table*}
\centering
\caption{Frequencies and amplitudes of SZ Lyn. V band data only are shown for Mount Abu and APT. The amplitude is in milli-magnitude. The errors of the amplitudes were obtained by Monte Carlo simulation using \textsc{Period04}. The 4$\sigma$ threshold limit for TESS is 0.08 which indicates by short horizontal line. }
\begin{tabular}{|r|r|r|r|r|r|r|r|r|}
\hline
& TESS & & Mount Abu & & APT & & WASP &\\
ID & Frequency & Amplitude & Frequency & Amplitude & Frequency & Amplitude & Frequency & Amplitude \\
& (c/d) & (mmg) & (c/d) & (mmg) & (c/d) & (mmg) & (c/d) & (mmg) \\
\hline
$f_{1}$ & 8.296 & 145.850$\pm$0.02 & 8.296 & 229.0$\pm$3.0 & 8.296 & 232.0$\pm$3.0 & 8.296 & 224.0$\pm$3.0\\
$2f_{1}$ & 16.592 & 53.320$\pm$0.01 & 16.590 & 76.0$\pm$3.0 & 16.592 & 82.0$\pm$3.0 & 16.593 & 76.0$\pm$3.0\\
$3f_{1}$ & 24.889 & 20.020$\pm$0.01 & 24.890 & 32.0$\pm$1.0 & 24.889 & 31.0$\pm$1.0 & 24.890 & 24.0$\pm$1.0\\
$4f_{1}$ & 33.186 & 8.220$\pm$0.01 & 33.186 & 11.0$\pm$1.0 & 33.185 & 13.0$\pm$1.0 & 33.185 & 15.0$\pm$1.0\\
$5f_{1}$ & 41.482 & 4.080$\pm$0.01 & 41.483 & 9.0$\pm$1.0 & 41.481 & 6.0$\pm$1.0 & 41.567 & 7.0$\pm$1.0\\
$6f_{1}$ & 49.779 & 2.420$\pm$0.01 & 49.849 & 3.0$\pm$1.0 & 49.777 & 4.0$\pm$1.0\\
$7f_{1}$ & 58.075 & 1.460$\pm$0.01 & 58.065 & 7.0$\pm$1.0 & 58.157 & 3.0$\pm$1.0\\
$8f_{1}$ & 66.372 & 0.800$\pm$0.01 & 66.365 & 4.0$\pm$1.0 & 66.339 & 3.0$\pm$1.0\\
$f_{2}$ & 14.535 & 0.800$\pm$0.01 & 14.221 & 6.0$\pm$1.0 \\
$9f_{1}$ & 74.668 & 0.470$\pm$0.01 & 73.120 & 3.0$\pm$1.0\\
$f_{2}+f_{1}$ & 22.830 & 0.309$\pm$0.008 & & \\
$f_{3}$ & 32.620 & 0.293$\pm$0.008 & & & \\
$10f_{1}$ & 82.965 & 0.277$\pm$0.008 & & \\
$f_{2}-f_{1}$ & 6.237 & 0.223$\pm$0.010 & & \\
$f_{4}$ & 4.584 & 0.164$\pm$0.007 & & & \\
$f_{3}+f_{1}$ & 40.888 & 0.157$\pm$0.010 & & & \\
$11f_{1}$ & 91.261 & 0.154$\pm$0.009 & & \\
$f_{3}-f_{1}$ & 24.308 & 0.144$\pm$0.007 & & & \\
$f_{1}-f_{4}$ & 3.703 & 0.143$\pm$0.009 & & & \\
$f_{1}+f_{4}$ & 12.880 & 0.120$\pm$0.006 & & & \\
$12f_{1}$ & 99.564 & 0.105$\pm$0.008 & & \\\cmidrule{1-3}
$13f_{1}$ & 107.824 & 0.068 & & \\
$14f_{1}$ & 116.104 & 0.042 & & \\
\hline
\end{tabular}
\label{frequencies of SZ Lyn}
\end{table*}

\begin{table}
\centering
\caption{Observed amplitudes ($A$ in magnitude) and phases ($\phi$ in radians) for the main frequency \textit{f$_{1}$}=8.296 c/d obtained from PERIOD04. Only \textit{f$_{1}$} is included in the table as this is the only independent frequency available in multi-band photometry of ground based data. The observed amplitude ratios in the middle row were used in Sec. \ref{mode}. The last row is phase differences in degrees. The phase zero is at 2457480.63391813 HJD.}
\begin{tabular}{ccccc}
\hline 
& U & B & V & R\\
\hline 
A & 0.275$\pm$0.003 & 0.290$\pm$0.004 & 0.229$\pm$0.003 & 0.171$\pm$0.002\\
$\phi$ & 0.834$\pm$0.003 & 0.805$\pm$0.003 & 0.803$\pm$0.003 \\
\hline
&$A_{U}/A_{B}$ & $A_{B}/A_{B}$ & $A_{V}/A_{B}$ & $A_{R}/A_{B}$ \\ 
&0.95$\pm$0.02 & 1.00 & 0.79$\pm0.02$ & 0.59$\pm0.02$ \\
\hline 
&$\phi_{U}-\phi_{B}$ & $\phi_{U}-\phi_{V}$ & &\\
&1.7$\pm0.2$ & 1.8$\pm0.2$ & &\\
\hline
\end{tabular}
\label{observational amplitudes}
\end{table}

\section{Amplitude Ratios}
\label{mode}
It is possible to determine the spherical degree $\textit{l}$ using the amplitude ratios and phase differences in different wavelength bands \citep[e.g.,][]{Balona1999,2010aste.book.....A}. The process is model-dependent because the amplitude and phase variations not only depend on the spherical harmonic degree \textit{l} \citep{Balona1999}, but also --to a lesser extent-- on the effective temperatures and gravities of the models as shown in \citet{Dupret2003}. In general, flux changes in pulsating stars originate from the temperature and gravity variations as a function of radius \citep{Garrido1990}. The mode identification needs to be done through a sequence of theoretical modeling of different combinations of parameters of the star. Theoretical aspects of flux change of the model star were discussed in detail with the consideration of non-adiabatic parameters and limb darkening effect by \citet{Watson1988} and \citet{Heynderickx1994}.

\begin{figure*}
\centering
\includegraphics[width=510pt,height=140pt]{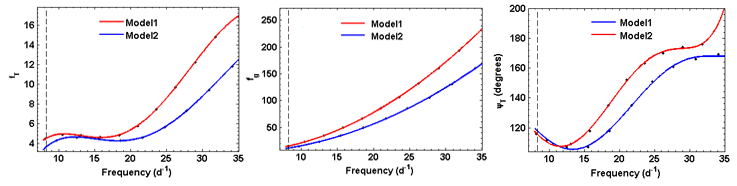}
\caption{$f_{T}$, $f_{g}$ and $\psi_T$ as a function of the pulsation frequency (d$^{-1}$) for the mode of degree \textit{l} = 0 for Model 1 and Model 2. The dots are the values generated by non-adiabatic code and continuous lines are the best fit polynomials. The dashed line indicates the observed frequency \textit{f$_{1}$}. The frequencies, \textit{f$_{2}$}, \textit{f$_{3}$} and \textit{f$_{4}$} are not marked since they were not identified in multi-band observation and hence could not include in Sec. \ref{mode}.}
\label{fig:EigenValues.png_figure}
\end{figure*}

The spherical degree can be identified by matching the observed amplitude ratios with the theoretically predicted amplitude ratios for the different stellar model atmospheres. The magnitude variation of a pulsating star, $\Delta m_\lambda$, at wavelength $\lambda$ with a spherical harmonic degree \textit{l} at a pulsating frequency $\omega$ can be expressed as in \citet{2010aste.book.....A}:
 
\begin{equation}
\Delta m_{\lambda} = A_0 P_{lm}(\cos\ i)b_{l\lambda}(T_1+T_2+T_3)e^{i\omega t}
\end{equation}
$T_{1} = (1-l)(l+2)$\\
$T_{2} = f_{T}(\alpha_{T\lambda}+\beta_{T\lambda})e^{-i\psi_T}$\\
$T_{3} = -f_{g}(\alpha_{g\lambda}+\beta_{g\lambda})$\\

\noindent where $P_{lm}$ is the associated Legendre function of degree $l$ and azimuthal number $m$, $A_{0}$ is related to the amplitude of oscillations of the photosphere, and $i$ is the inclination angle between the stellar axis and the direction towards the observer. For the derivation and the details of the components in Eq. 1, refer to \citet{Watson1988}, \citet{Heynderickx1994}, \citet{Balona1999} and \citet{Dupret2003}. The component $T_1$ is the contribution of the magnitude variation due to the different pulsation modes. $T_2$ is the temperature dependent component of the magnitude which consists of $f_T$, the amplitude of temperature variation function relative to the normalized radial displacement at the photosphere and $\psi_T$, the phase difference between maximum temperature and maximum radial displacement. $T_3$ represents gravity variation where $f_g$ is the amplitude of gravity variation function to the normalized radial displacement at the photosphere \citep{Dupret2003}. The determination of $f_T$ is highly dependent on the non-adiabatic parameter R \citep{garrido2000photometric}. The other two parameters, $\psi_T$ and $f_g$ are also unknown for any stellar model. The method of mode identification therefore depends on the correct combination of these parameters which were done earlier by different techniques, approximating the observations to the theoretical models for reasonable ranges of $f_T$ and $\psi_T$ \citep{Balona1999}. The value of R was estimated by \citet{Heynderickx1994} by adjusting it for best fit of the observed light curves, while \citet{Garrido1990} redefined it as a range of interest keeping R and $\psi_T$ as free parameters in the range [0.25\ $-$ \ 1] and [ $90^{\circ}-135^{\circ}$] respectively for $\delta$ Scuti stars. \citet{Dupret2003} introduced Time Dependent Convection (TDC) non-adiabatic treatment of the model atmosphere and proposed a method of determination of amplitudes $f_T$, $f_g$ of the  eigenfunctions and the phase $\psi_T$ for different degree \textit{l} for a range of effective temperatures T$_{\rm eff}$, $\log(g)$ and mass of a star. In this work, we used the outputs of TDC  non-adiabatic computations provided by M. A. Dupret (private communication) for SZ Lyn. The input parameters T$_{\rm eff}$=7540 K, $\log(g)$=3.88 \citep{Langford1976} and M=1.57 M$_\odot$ \citep{Fernley1983}  with the observational uncertainties (see Table \ref{tab:params}) were passed to the non-adiabatic code to generate the pulsation models which give frequencies close to the observed main frequency of 8.296 d$^{-1}$. Details of this iterative process can be found in \citet{dupret2005}. Two models which found very close to our observation results are given in Table \ref{theoreticalmodels}. The output of the non-adiabatic code also provides the amplitudes of $f_T$, $f_g$ and phase angle $\psi_T$ for frequencies for different degree \textit{l}. Fig. \ref{fig:EigenValues.png_figure} represents the distribution of $f_T$, $f_g$ and $\psi_T$ with the model frequencies generated by non-adiabatic code. The interpolations of the best fit polynomials determine the $f_T$, $f_g$ and $\psi_T$ for the observed frequency \textit{f$_{1}$}. We have shown only the $l=0$ mode for the present purpose in Fig. \ref{fig:EigenValues.png_figure} . \\

\begin{table}
\centering
\caption{Best-fitting DTC non-adiabatic pulsation models for the observed frequencies of SZ Lyn.}
\begin{tabular}{ccc}
\hline
& \textbf{Model 1} \\
$M/M_{\odot}$=2.00 & $T_{\rm eff}$=7522 K & $\log(L/L_{\odot}$)=1.425\\
$\log(g)$=3.77 & $R/R_{\odot}$ = 3.00 & age(yr)=$1.0\times10^{9}$\\
X=0.72 & Z=0.014 & $\alpha_{\rm MLT}$ = 1.7\\
\hline
& \textbf{Model 2} \\
$M/M_{\odot}$=1.90 & $T_{\rm eff}$=7557 K & $\log(L/L_{\odot}$)=1.322\\
$\log(g)$=3.86 & $R/R_{\odot}$ = 2.68 & age(yr)=$1.1\times10^{9}$\\
X=0.72 & Z=0.014 & $\alpha_{\rm MLT}$ = 1.7\\
\hline
\end{tabular}
\label{theoreticalmodels}
\end{table}  
    
\begin{figure}
\centering
\includegraphics[width=230pt,height=368pt]{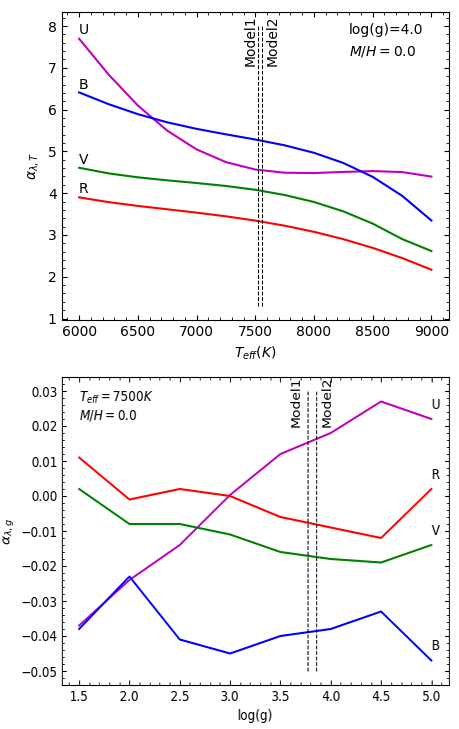}
\caption{Flux derivative as a function of temperature (upper panel) and as a function of gravity (lower panel) generated by AlphaTg code. The two dashed line indicates the two models proposed for SZ Lyn in Table \ref{theoreticalmodels}. The derivatives taken at these two models were used for the calculation of theoretical amplitudes.}
\label{fig:alphaTvslogT.png_figure}
\end{figure}

For the partial derivatives of the monochromatic flux, $\alpha_{T\lambda}$ (rate of change in flux due to temperature) and $\alpha_{g\lambda}$ (rate of change in flux due to gravity) (Eq. 1), we used \textsc{ATLAS9} model atmospheres and fluxes by \citet{1993KurCD}, \citet{Castelli2003}. Due to the complexity of lookup tables in all UBVR bands of Kurucz grid for the computation of $\alpha_{T\lambda}$ and $\alpha_{g\lambda}$, we produced a code, AlphaTg, to readout the model fluxes from the grid for the desired range of effective temperature and $\log(g)$ with a step size of 250 K and 0.5 respectively. Since the non-adiabatic parameters of $f_T$, $f_g$ and $\psi_T$ were computed for the solar metallicity of $Z=0.014$, we used the solar metal abundance of $[M/H]=0.0$ model from the Kurucz grid for the partial derivatives. Furthermore, \citet{garrido2000photometric} has shown that the flux derivatives have no significant change with the metallicity except in blue band. The variation of flux derivatives in Fig. \ref{fig:alphaTvslogT.png_figure} also provides similar results as the relative variation is much higher for shorter wavelengths.  \citet{Balona1999} pointed out that amplitude discrimination is more effective in shorter wavelength bands because of this variation in flux derivatives. All the flux models were computed with the turbulence velocity fixed at 2 km/s and a mixing length parameter $(\alpha_{\rm MLT})$ of 1.25. \citet{breger1998delta} showed the mixing length parameter, $\alpha_{\rm MLT}$, is in between 1 and 2 for $\delta$ Scuti stars and less effective on hot stars while later according to \citet{Bowman2018aap} the $\alpha_{\rm MLT}$ was revised as close to 2 for $\delta$ Scuti stars. Hence the flux derivatives were calculated for $\alpha_{\rm MLT}$ = 1.25 which is the highest of two options, 1.25 and 0.5 given in \textsc{ATLAS9} models. The partial derivatives, $\alpha_{T\lambda}$ and $\alpha_{g\lambda}$ computed within the observational error box of SZ Lyn using AlphaTg code are shown in Fig. \ref{fig:alphaTvslogT.png_figure}. It is clear that the temperature contribution ($\alpha_{T\lambda}$) to the magnitude variation and hence to the theoretical amplitudes is much more effective than the gravity component ($\alpha_{g\lambda}$), as seen in Fig. \ref{fig:alphaTvslogT.png_figure}. 

Limb darkening effect was added to the theoretical amplitude by computing limb darkening integral using the simple linear limb darkening law given by \citet{Claret2003}  
 
\begin{equation}
b_{l\lambda} = \int_{0}^{1} \mu I(\mu) P_l d\mu
\end{equation}
$\frac{I(\mu)}{I(1)} = 1-u(1-\mu)$ \\

\noindent where $u$ is the limb darkening coefficient.\\

\begin{figure}
\centering
\includegraphics[width=230pt,height=368pt]{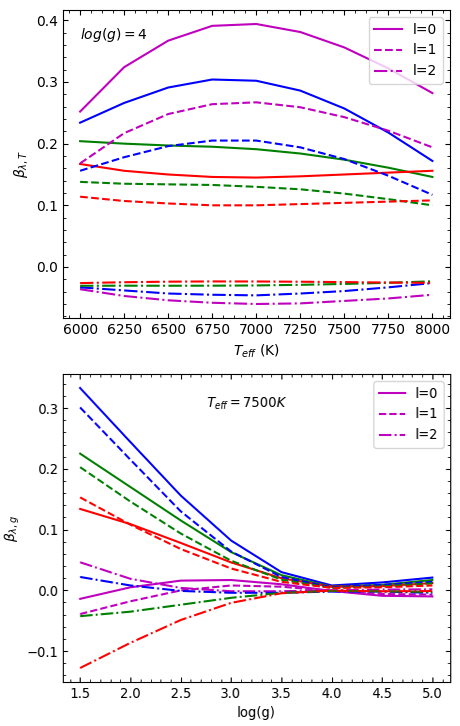}
\caption{Limb darkening derivatives. The upper panel shows the variation of limb darkening derivative with temperature for all three modes of $\textit{l}$ and for UBVR colour bands for $\log(g)=4$. The colours, Violet, Blue, Green and Red represent the UBVR bands respectively. Lower panel shows limb darkening derivatives with gravity for $l$ and UBVR for T$_{\rm eff}$ = 7500 K.}
\label{fig:Beta_T.png_figure}
\end{figure}

\begin{figure}
\includegraphics[width=250pt]{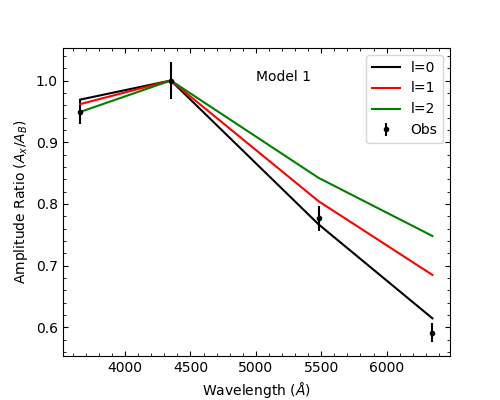}
\caption{The comparison of the observed amplitude ratios of the frequency \textit{f$_{1}$} with the computed amplitude ratios for the Model 1 for the frequency $f_1$. Dots with error bars are the observed amplitude ratios and the lines are the theoretical predictions for spherical degree, \textit{l}.} 
\label{fig:AmplitudeRatios1.png_figure}
\end{figure}

\begin{figure}
\includegraphics[width=250pt]{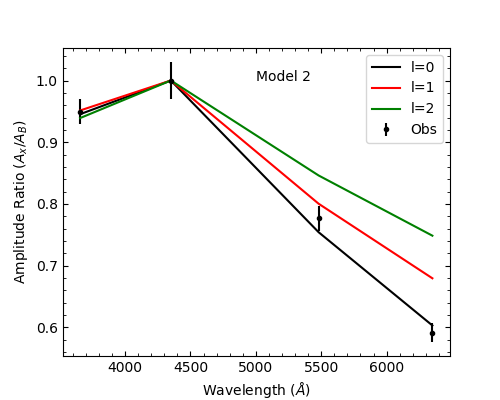}
\caption{The comparison of the observed amplitude ratios of the frequency \textit{f$_{1}$} with the computed amplitude ratios for the Model 2 for the frequency $f_1$. Dots with error bars are the observed amplitude ratios and the lines are the theoretical predictions for spherical degree, \textit{l}.} 
\label{fig:AmplitudeRatios2.png_figure}
\end{figure}

The limb darkening integral $\textit{b}_{l\lambda}$ depends on the mode of the oscillation $\textit{l}$. From the grids of limb darkening coefficients provided by \citet{Claret2003}, the limb darkening integrals were computed using AlphaTg code for UBVR bands and hence obtained derivatives, $\beta_{T\lambda}$ and $\beta_{g\lambda}$, for three spherical degrees of $l$ ($l$=0,1,2) in the range of effective temperatures and gravity of SZ Lyn. The results are shown in Fig. \ref{fig:Beta_T.png_figure}. The limb darkening derivatives of temperature in upper panel of Fig. \ref{fig:Beta_T.png_figure} are more sensitive to \textit{l} = 0 degree as well as for lower wavelengths while derivatives are less effective for \textit{l} = 2. Besides, the gravity counterpart of limb darkening derivative (lower panel, Fig. \ref{fig:Beta_T.png_figure}) compared to the temperature is very much less particularly in the range of $\log(g)$ of SZ Lyn. The effect of gravity change is less effective for mode discrimination than that of temperature, suggesting that the effect of gravity on limb darkening is negligible compared to the effect of temperature. This effect is previously shown by \citet{garrido2000photometric}. However, we consider limb darkening derivatives due to gravity, even though negligible, in our calculations of theoretical amplitudes because of the model parameters of SZ Lyn in Table \ref{theoreticalmodels} are very close to each other. Based on the computations, The theoretical amplitude ratios of $l$ = 0, 1 and 2 for Model 1 and Model 2 given in Table \ref{theoreticalmodels} were calculated and subsequently compared with the observed amplitude ratios given in the Table \ref{observational amplitudes} for the frequency \textit{f$_{1}$}. The graphical representation of the comparison of amplitude ratios for Model 1 and Model 2 are shown in Fig. \ref{fig:AmplitudeRatios1.png_figure} and Fig. \ref{fig:AmplitudeRatios2.png_figure}, respectively. Since observations are scattered and the two model parameters are very close to each other, in order to better assess the appropriateness of the different values for $l$ as well as to converge for the best model, we computed the $\chi^2$ function in Eq. 3 \citep{daszynska2009} for different degree \textit{l} for the two models:\ 

\begin{equation}
\chi^2(l)=\sum_{j=1}^{\rm filters}\Big(\frac{A_{j,\rm th}/A_{\rm ref,th}-A_{j,\rm obs}/A_{\rm ref,obs}}{\sigma_{j,\rm obs}}\Big)^2
\end{equation}

\begin{figure}
\centering
\includegraphics[width=200pt]{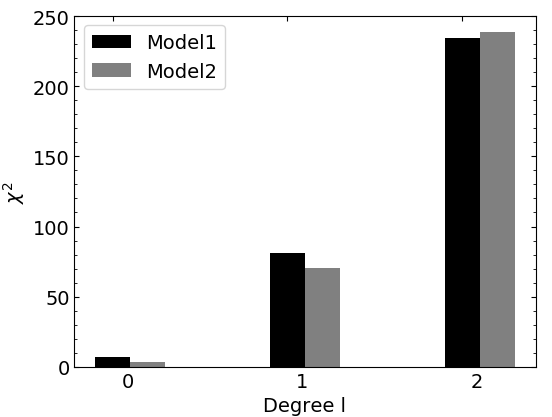}
\caption{$\chi^2$ minimization of the theoretical and observational amplitudes for the frequency \textit{f$_1$} against the spherical degree \textit{l} for Model 1 and Model 2.}  
\label{fig:Chi1.png_figure}
\end{figure}

From the $\chi^2$ minimization shown in Fig. \ref{fig:Chi1.png_figure} we can conclude that the frequency \textit{f$_{1}$} is better fitted to $l=0$ degree of Model 2 and therefore Model 2 is more appropriate to represent the physical properties of SZ Lyn.       

\section{Physical Parameters of SZ Lyn}
\label{para}
We have tried to refine the physical parameters of SZ Lyn in this asteroseismic investigation. Although the fundamental radial mode, \textit{f$_{1}$}, is clearly defined in early literature, we used asteroseismic approach with state-of-art TDC non adiabatic models to determine the spherical degree of the frequency \textit{f$_{1}$} through the mode identification method in Sec. \ref{mode}. The advantage of this approach is the ability to investigate stellar parameters through different models of different pulsation codes as well as the possibility of perturbation of parameters over the acceptable ranges within the models; hence we can obtain the best set of physical parameters.   

\subsection{HELAS model oscillations}

Furthermore, we considered HELAS\footnote{\url{http://helas.astro.uni.wroc.pl/deliverables.php?active=opalmodel&lang=en}} pulsation models with the observed frequencies of SZ Lyn. We searched the HELAS pulsation models produced by the linear non-adiabatic 
code \citep{pamyatnykh1999,dziembowski1977} using OP opacities \citet{seaton2005} and AGSS09 chemical composition \citet{Asplund2009} for Z=0.015 and Z=0.02, from ZAMS to early post-main sequence stages, and for radial and non-radial oscillations.

The HELAS pulsation models of mass range from M=$1.8$ M$_\odot$ to M=$2.0$ M$_\odot$ with the step size of M=$0.1$ M$_\odot$ are consistent with the observed radial mode \textit{f$_{1}$}, 8.296 d$^{-1}$ (see Table \ref{HELAS models}). 
In addition, the other two frequencies, \textit{f$_{2}$}, \textit{f$_{3}$} are also overlapping with the model frequencies of \textit{l}=2 mode of the model of mass M=$2.0$ M$_\odot$. Therefore, our identification of \textit{f$_{2}$} and \textit{f$_{3}$} as non-radial modes in both our frequency analysis and our pulsation constant calculations is strengthened by these model calculations, which have non-radial modes at frequencies close to the observed values for \textit{f$_{2}$} and \textit{f$_{3}$}. Although this is not sufficient for a solid conclusion of the spherical degree of \textit{f$_{2}$} and \textit{f$_{3}$} as \textit{l}=2, we can still conclude that they likely are non-radial modes. From the models with predicted fundamental radial node \textit{n}=1, \textit{l}=0 given in Table \ref{HELAS models}, the model M=1.9 M$_\odot$ is in line with the parameters of the TDC non-adiabatic Model 2 approximated by amplitude ratio method. 

Therefore, HELAS models also provide evidence favouring a star of mass M$\approx$1.9 M$_\odot$ which oscillates in radial fundamental mode close to 8.296 d$^{-1}$ as well as having some non-radial modes close to the observed frequencies. Hence, considering all the results of this section, we infer that the mass of SZ Lyn should be within 1.8 M$_\odot$ $\le$ M $\le$ 2.0 M$_\odot$.\\ 

\begin{table}
\centering
\caption{HELAS fundamental radial pulsation models (\textit{n}=1,\textit{l}=0) close to SZ Lyn.}
\begin{tabular}{lllllll}
\hline
Model & Mass  & T$_{\rm eff}$  & log (L/L$_\odot$) & $\log(g)$ & Z & Freq.   \\
 & (M$_\odot$) & (K) & & & & (d$^{-1}$) \\ 
\hline
H1 & 1.8 & 7246 & 1.324 & 3.76 & 0.015 & 8.230 \\
H2 & 1.9 & 7574 & 1.414 & 3.77 & 0.015 & 8.269 \\
H3 & 2.0 & 7892 & 1.501 & 3.78 & 0.015 & 8.274 \\
H4 & 1.9 & 7205 & 1.326 & 3.77 & 0.02 & 8.250 \\
H5 & 2.0 & 7518 & 1.413 & 3.78 & 0.02 & 8.299 \\
\hline
\end{tabular}
\label{HELAS models}
\end{table}

\subsection{Evolutionary status of SZ Lyn}

We have computed evolutionary sequences using version 11701 of the MESA stellar evolution code \citep{Paxton2011,Paxton2013,Paxton2015,Paxton2018,Paxton2019} for stars with masses ranging from $1.70$ M$_\odot$ to $2.0$ M$_\odot$ in steps of $0.1$ M$_\odot$.
The evolutionary models were computed for the standard solar composition of (\citet{Asplund2009}, hereinafter AGSS09) scaled to $Z=0.01$ and $Z=0.02$ to bracket the reasonable values of metallicity for SZ Lyn. All the sequences use OPAL opacities \citep{Iglesias1996} and a single fixed value for the mixing length parameter of $\alpha_{\rm MLT}=2.0$. The differences between OPAL and the OP opacities used in HELAS models are rather small in the range of effective temperature of SZ Lyn \citep{seaton2004comparison}, thus allowing a comparison between the two sets. We have adopted the model of exponential decay \citep{Herwig2000} to account for core overshooting. In this model, raising convective elements exponentially decay after crossing the boundary marked by Schwarzschild's convective criterion:

\begin{equation}
D_{\rm ov}=D_{{\rm conv},0} \exp{\left(-\frac{2z}{f_{\rm ov}H_{{\rm P},0}}\right)}
\end{equation}

\noindent where $D_{{\rm conv},0}$ is the diffusion coefficient near to the convection boundary, $H_{{\rm P},0}$ the pressure scale height (both determined according to the MLT). In our sequences, we have used three values of the parameter $f_{\rm ov}$: 0.00, 0.007, and 0.014 to model several extents of overshooting, from none ($f_{\rm ov}=0.00$) to a significant overshooting ($f_{\rm ov}=0.014$).

The nuclear network is \textit{pp\_and\_cno\_extras.net}, that includes 25 isotopes ranging from $^1$H to $^{24}$Mg. We have neither considered mass-loss nor rotation for the evolutionary sequences.\\

\begin{figure}
\centering
\includegraphics[width=\columnwidth]{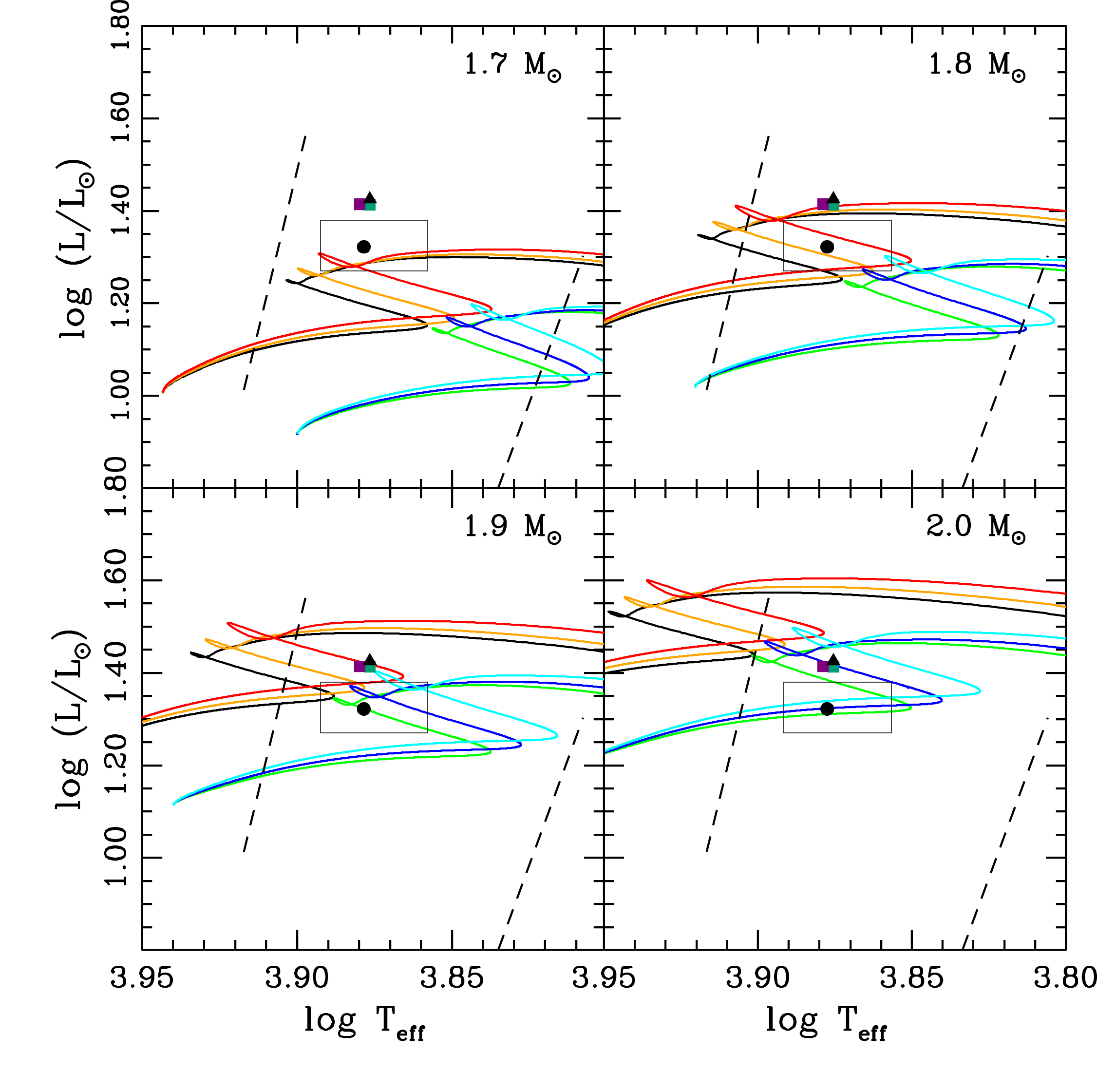}
\caption {Main and post-main sequence evolutionary tracks of stars in the range 1.7-2.0 M$_\odot$ for the scaled AGSS09 chemical composition $Z=0.01$ (upper set of tracks in each panel) and $Z=0.02$ (lower set of tracks in each panel) for $\alpha_{\rm MLT}=2$. Curves of green and black colours correspond to $f_{\rm ov}=0.00$, dark blue and orange to $f_{\rm ov}=0.007$, and light blue and red to $f_{\rm ov}=0.014$. The black triangle and dot represent TDC Model 1 and Model 2, respectively (Table \ref{theoreticalmodels}). The purple and green squares represent Models H2 and H5 from Table \ref{HELAS models}, respectively. The rectangle shows observation error box. Instability strip is shown in dashed lines.}
\label{fig:EvolTracks_logL.png_figure}
\end{figure}

The full set of evolutionary sequences is shown in Figure \ref{fig:EvolTracks_logL.png_figure}. The tracks for low metallicity are of higher luminosity, and the ones for higher metallicity are of lower luminosity. Increasing core overshooting ($f_{\rm ov}=0.00, 0.007, 0.014$) causes an increase in the duration of the main sequence, as well as an evolution at a slightly higher luminosity. Dashed lines show the limits of the instability strip \citep{breger1998}. An analysis of this figure shows that, given the ample range of parameters studied, there is not a single match for the TDC non-adiabatic pulsation models (Model 1 is marked as a triangle, Model 2 as a circle), and that reasonable values of the mass can range from 1.7 to 2.0 solar masses (1.7 to 1.9 M$_\odot$ if $Z=0.01$, and 1.8 to 2.0 M$_\odot$ if $Z=0.02$). During the main sequence evolution of stars with M$\geq$1.2 M$_\odot$ a convective core appears, but its extension has not been still accurately determined (see, for example, the results on some of the stars analysed in \citet{Deheu2016}). As expected, the effect of increasing overshooting is to extend the duration of the main sequence, as a larger amount of H becomes available. For our purposes this effect obscures the current evolutionary state of SZ Lyn, which could be at the last phases of the main sequence, or at the first stages of the post-main sequence. Nevertheless, the evolutionary sequences indicate that the likely mass of SZ Lyn might be in the range 1.7--2.0 M$_\odot$, significantly larger than the old value of 1.57 M$_\odot$ found by \citet{Fernley1983}. Further observational refinements in the metallicity of the star would allow improvements in the mass inference.

\subsection{Pulsation constant}
\label{pulsation}
The pulsation constant ($Q$) can be determined for the frequency \textit{$f_1$} using the Eq. 5 given by \citet{1990DSSN} for the two models mentioned in Table \ref{theoreticalmodels}.\ 

\begin{equation}
\log(Q)=\log(P)+\frac{1}{2}\ \log(g)+\frac{1}{10}\ M_{\rm bol} + \log(T_{\rm eff}) - 6.454
\end{equation}

The bolometric magnitude M$_{\rm bol}$ of SZ Lyn is estimated for  two models using the values found in Table \ref{theoreticalmodels}. We take for the solar bolometric absolute magnitude the standard value of M$_{\odot{\rm bol}}$=4.74 \citep{Bessell, Torres} and solar luminosity L$_{\odot}=3.85\times 10^{33}$ erg s$^{-1}$. The pulsation constant $Q$ is then found to be 0.032 for Model 1 and 0.038 for Model 2. The stellar parallax of SZ Lyn obtained by {\it Gaia} mission provides an absolute magnitude M$_V$ of +1.14. and the bolometric correction (BC) is -0.094 as given by \cite{2000Cox}, thus resulting in a bolometric absolute magnitude for SZ Lyn of M$_{\rm bol}$ +1.05. This results in an alternative set of $Q$ values of 0.031 and 0.035 for Model 1 and Model 2 respectively. The pulsation constant for fundamental radial \textit{p}-modes in $\delta$ Scuti stars is in the range of $0.022 \leq Q \leq 0.033$ \citep{Breger1975b} and an even narrower range of $0.0327 \leq Q \leq 0.0332$ for the fundamental radial mode pulsation \citep{1981Apj}.

The calculations show that the $Q$ value for Model 1 is within the allowed range given by \citet{Breger1975b} and close to the lower limit of the range given by \citet{1981Apj}, and this gives reinforces our confidence that the frequency \textit{f$_{1}$} is the radial fundamental mode of stellar pulsation. On the other hand, the $Q$ value of radial fundamental mode of Model 2 parameters is slightly deviated from the standard intervals.

\subsection{Mean density}           
As we have only determined a single radial mode without any strongly confirmed overtones, we cannot use the period ratios to determine the average density of the star. Nevertheless, we could estimate the average density of SZ Lyn using the $Q$ values, calculated in Sec. \ref{pulsation}, using the relation:

\begin{equation}
P=Q\;(\rho/\rho_{\odot})^{-1/2} 
\end{equation}
where $P$ is the period of the fundamental oscillation in days and $Q$ is the pulsation constant. The average density of the Sun, $\rho_{\odot}$= 1.4103 g cm$^{-3}$, and predicted $Q$ values of 0.031 and 0.035 in Sec. \ref{pulsation} results in an average density of $\rho$=0.0933 gcm$^{-3}$ and $\rho$=0.1189 gcm$^{-3}$ for the Model 1 and Model 2 respectively. Though the fundamental period P is been determined with a high accuracy, the uncertainty of Q is obtained from a propagation of errors of Eq. 5 which could not be determined accurately due to the non-availability of model uncertainties of T$_{\rm eff}$ and $\log(g)$. Therefore, the determination of average density using Q values calculated in Sec. \ref{pulsation} is unreliable. 
Instead, with the accurate fundamental period of SZ Lyn, $P=0.120526$ d and for more narrow range of $Q$ values $0.0327 \leq Q \leq 0.0332$ \citep{1981Apj}, we find the average density and error as $\rho=0.1054\pm0.0016$ gcm$^{-3}$ from the upper and lower limits of the above range.    \citet{suarez2014} computed the mean densities of $\delta$ Scuti stars using their virtual observatory tool, TOUCAN, and found that the fundamental frequency range of 95 - 113 $\mu$Hz (0.121832 - 0.102425 d) has relative densities ($\rho/\rho_{\odot}$) less than 0.11. Hence we can conclude that the density of $\rho=0.1054\pm0.0016$ gcm$^{-3}$ is an appropriate value for SZ Lyn.  

\begin{figure*}
\centering
\includegraphics[width=500pt]{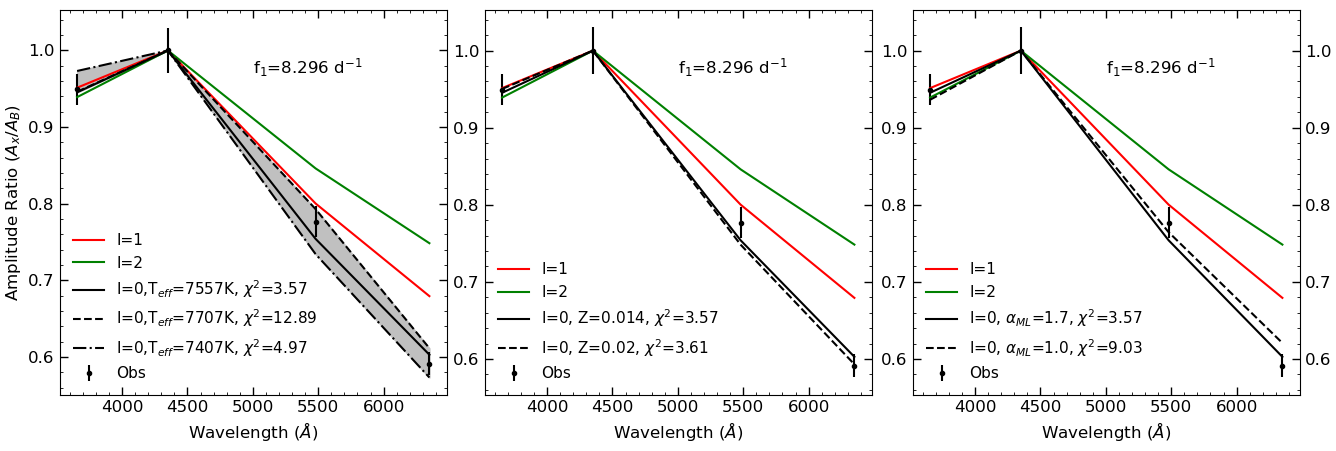}
\caption{Perturbation of input parameters T$_{\rm eff}$, chemical composition (Z) and mixing length ($\alpha_{\rm MLT}$) on \textit{l}=0 of Model 2. Left; Deviation of theoretical amplitude ratios at \textit{l}=0 with observational error of $\pm$150 K at constant Z=0.014 and $\alpha_{\rm MLT}$=1.7. Middle; Deviations of amplitudes with Z at T$_{\rm eff}$=7557 K and and $\alpha_{\rm MLT}$=1.7. Right; Deviation of amplitudes with $\alpha_{\rm MLT}$ at T$_{\rm eff}$=7557 K and Z=0.014. The $\chi^2$ minimization with the observation for each deviation is included.}
\label{fig:uncertanity.png_figure}
\end{figure*}

\section{Discussion}
\label{discuss}
SZ Lyn has been extensively studied by several authors who have analysed their results considering both its binary nature as well as its pulsating nature. Though its binary orbital characteristics have been studied in detail, a comprehensive asteroseismic study of SZ Lyn was lacking so far. We reconsider SZ Lyn giving special attention to recovering  more frequencies apart from the main frequency, and determining the degree of the oscillation. With this information, we have tried to refine its stellar parameters.\

We confirmed 23 frequencies, 14 of which are multiples of the fundamental frequency \textit{f$_{1}$}. \citet{Gazeas2004} also confirmed the frequency \textit{f$_{1}$} and its first two harmonics, and we discovered 11 new harmonics in SZ Lyn.  Therefore to confirm the radial mode, we adopted the amplitude ratio method as it provides a consistent result with the observed amplitudes. In addition to the amplitudes, $\delta$ Scuti stars show wavelength dependence on the phase. A method proposed by \citet{Garrido1990} showed that mode can be determined by plotting amplitude ratio versus phase difference for two colors. This method was performed for Str\"omgren photometric system taking two bands each time and plotting the so-called area of interest for a broad range of parameters of $\psi_{T}$ and non-adiabaticity R. Although the phase difference method is more appropriate for $\delta$ Scuti stars to determine the degree of the modes, we did not represent it graphically because of the availability of one frequency (\textit{f$_{1}$}) in multi-band photometry and there are no other frequencies to compare. But the phase difference of the frequency \textit{f$_{1}$} shown in Table \ref{observational amplitudes}, $\phi_{\rm U}-\phi_{\rm B}$ and $\phi_{\rm U}-\phi_{\rm V}$ both are greater than zero \citep{2000Balt,2000ASPC}, thus further confirming that f$_{1}$ corresponds to a radial mode of \textit{l}=0. Furthermore, the TESS observations clearly revealed the presence of independent frequencies \textit{f$_{2}$}=14.535 d$^{-1}$, \textit{f$_{3}$}=32.620 d$^{-1}$ and \textit{f$_{4}$}=4.584 d$^{-1}$. Unfortunately these frequencies could not be recovered in the multi-band ground based data. Therefore, we could not perform the amplitude ratio method to determine the spherical degree \textit{l}. In some cases, mode identification can be done without multi-band photometry as more often the space data has shown rotational splitting and period spacing patterns in the frequency spectra \citep{aerts2019,2010aste.book.....A}.

Provided that the frequency f$_{2}$ is nonradial, in principle, the effect of rotational splitting should place side lobes centered at f$_{2}$. We estimated the first-order effect of rotational splitting using the measured upper limit of $v \sin i < 40$\,km s$^{-1}$ \citep{1985PASP...97..715M}. Given that this is also a general limit to the projected rotational velocity of high-amplitude $\delta$ Scuti stars \citep{2000ASPCBreger}, we can assume this as an intrinsic limit to $v_{\rm rot}$. Using R$_{*}$=2.86 R$_{\odot}$ from our Model 2, the upper limit of first order rotational splitting for nonradial modes was estimated with a negligible Ledoux constant, C$_{n.l}$ \citep{ledoux1951}. This upper limit amounts to 0.276 d$^{-1}$. We were not able to find signatures of a recurring frequency splitting within this limit around f$_{2}$ in SZ Lyn. The absence of rotational splitting and period spacings in the frequency spectra of SZ Lyn shows the importance of continuous multi-band ground based observations to determine \textit{l} using amplitude ratios and phase differences. However. we were able to conclude that \textit{f$_{2}$} and \textit{f$_{3}$} are non-radial modes through the period ratio method and pulsation constant calculations. In addition, the HELAS pulsation models give some evidences that \textit{f$_{2}$} and \textit{f$_{3}$}  are close to \textit{l}=2 mode but we refrain from confirming the mode of \textit{f$_{2}$} and \textit{f$_{3}$} as further investigations are needed. The discovery of frequency \textit{f$_{4}$} is remarkable because it is identified as medium-order \textit{g}-mode pulsation which is rare in $\delta$ Scuti stars.\

As a consequence of the determination of independent frequencies and their modes, this investigation is used to refine the stellar parameters of SZ Lyn. In the process of determining the degree of the mode \textit{l} of the frequency \textit{f$_{1}$} using amplitude ratio method, it is possible to narrow down the stellar parameters of SZ Lyn. We produced pulsation models using TDC non adiabatic models \citep{Dupret2003} as well as linear non adiabatic models \citep{Pamyatnykh, dziembowski1977}. The introduction of state-of-art TDC non adiabatic models was an important step forward in the modelling of $\delta$ Scuti stars \citep{Dupret2003,dupret2005}. \citet{murphy2012} gave a successful first application of these models to an individual $\delta$ Scuti star observed by Kepler. Finally, evolutionary sequences were computed using MESA and overlapped the models to provide a further insight on the stellar parameters. Although the determination of stellar parameters through the combination of different models is more accurate, the inconsistencies with models clearly stand out. Therefore, the behaviour of the models was considered by perturbing the input parameters. \\

We searched the best model for SZ Lyn centering the error box at T$_{\rm eff}$=7500 K and $\log(g)=4.0$ and stepping in T$_{\rm eff}$ ($\pm$250 K) and in $\log(g)$ ($\pm$0.5) in \textsc{ATLAS9}. The best fit polynomials (minimum RMSE) of AlphaTg code determine the flux and limb darkening derivatives for any T$_{\rm eff}$ and $\log(g)$ so that the error contribution from flux and limb darkening derivatives to the theoretical amplitudes is minimized. However, the perturbation of T$_{\rm eff}$ on eigenvalues $f_{T}$ and $\psi_{T}$ is significant and hence change the theoretical amplitudes considerably \citep{dupret2005}. The temperature component (T$_2$) has a major contribution in theoretical amplitudes. Thus the best model was perturbed by the observational uncertainty of $\pm$150 K. This uncertainty propagates to the non adiabatic parameters $f_T$ and $\psi_T$ and thus deviated the amplitudes of \textit{l}=0 as shown in the left panel of Fig. \ref{fig:uncertanity.png_figure}. \citet{dupret2005} investigated the dependency of non adiabatic parameters $f_T$ and $\psi_T$ for effective temperature range of $\delta$ Scuti stars and showed that in the lower effective temperature region ($\approx$ 7250 - 6500 K) both $f_T$ and $\psi_T$ drop significantly. We performed the extrapolation to the lower temperature region and observed the amplitude ratios of all pass bands were further reduced and hence deviated more from the observations. The temperature perturbation of Model 2 shown in Fig. \ref{fig:uncertanity.png_figure} revealed that the $\chi^2$ value is minimized around 7550 K and deviated from the minimization at higher and lower T$_{\rm eff}$. This temperature approximation is consistent with linear non adiabatic models, H2 and H5, in Table \ref{HELAS models}.

By means of linear regression \citet{bowman2018mnras} obtained new statistics for $\delta$ Scuti stars using short cadence Kepler observations. In this analysis the stars are classified in three categories with respect to $\log(g)$ and 3.5$\le\log(g)\le$4.0 is defined as Middle Age Main Sequence (MAMS). By means of the linear regression coefficients, the T$_{\rm eff}$ was estimated for the observed frequency f$_1$ = 8.296 d$^{-1}$ as 7620 K for revised KIC values from \citet{huber2014} and as 7469 K for \citet{brown2011} KIC values. Though the specific T$_{\rm eff}$ is not possible to be given with model uncertainties, it is rather enough evidences supporting that the temperature of SZ Lyn converge to 7500 K$\le$T$_{\rm eff}$ $\le$7800 K resulting in the conclusion that SZ Lyn is more close to blue edge than thought before.\\ 

Similarly we investigated the model dependency on Z and $\alpha_{\rm MLT}$ in middle and right panels of Fig. \ref{fig:uncertanity.png_figure} respectively. The higher the metallicity (Z), the more efficient the $\kappa$ mechanism in lowering the luminosity in the driving region. Hence the amplitude of the luminosity variation and the local effective temperature variation $f_T$ at photosphere are small. In addition, The phase difference between the local effective temperature and radial displacement $\psi_T$ is close to 180$^{\circ}$ \citep{Dupret2003}. These individual variations were accounted in the computation of amplitudes in pass bands and overall effect on the amplitudes can be seen in the Fig. \ref{fig:uncertanity.png_figure}. The left panel of Fig. \ref{fig:uncertanity.png_figure} is the dependency of Model 2 on the mixing length parameter ($\alpha_{\rm MLT}$). The deviations of $f_T$ and $\psi_T$ for $\alpha_{\rm MLT}$ = 1.0 were taken from \citet{dupret2005}. $\alpha_{\rm MLT}$ = 1.0 is somewhat low for $\delta$ Scuti stars; a value of 1.8 to 2.0 provides a reasonably good agreement between theoretical models and observations of $\delta$ Scuti stars \citep{bowman2018mnras}. This is close to our choice of $\alpha_{\rm MLT}=1.7$ in Fig. \ref{fig:uncertanity.png_figure}. However, it is important to have more individual observations of $\delta$ Scuti stars, with a similar analysis, for a better understanding of the behaviour of these parameters.\

In this process the pulsation models of SZ Lyn were evaluated with an appropriate set of evolutionary sequences with different stellar parameters. We computed evolutionary sequences with MESA to obtain an independent value for the mass of SZ Lyn. Placing Models 1 and 2 in the HR diagram allowed us to compare with the evolutionary tracks of stars between 1.7--2.0 M$_\odot$, $Z=0.01-0.02$, and different values for overshooting. We believe that these ranges of mass, metallicity and overshooting encompass the likely values for SZ Lyn. The results show that both models are compatible with stars in these ranges at the last phases of the central H burning, or at the beginning of shell burning. Nevertheless, without a better determination of metallicity --and of the still harder to determine overshooting-- we cannot refine the mass determination beyond this 1.7--2.0 solar masses range. This range is in reasonable agreement with HELAS's models which converged to 2 M$_{\odot}$ model with the observation. \

\begin{table}
\centering
\caption{Best model of SZ Lyn. The $\log(g)$ is the average of all the pulsation models.}
\begin{tabular}{lcc}
\hline
Physical Property & Value \\
\hline
$T_{\rm eff}$  & 7500 K$\le$T$_{\rm eff}$$\le$ 7800 K \\
$\log (g)$  & 3.81$\pm$0.06 \\
Mass (M) & 1.7-2.0 M$_{\odot}$ \\
Radius (R) & 2.68 R$_{\odot}$ \\
Metallicity (Z) & 0.014 - 0.020 \\
Mean density ($\rho$) & 0.1054$\pm$0.0016 g cm$^{-3}$ \\
\hline
\end{tabular}
\label{final parameters}
\end{table}

\section{Conclusions}
\label{conclusion}

A comprehensive analysis of ground and space based photometry of the $\delta $Scuti star SZ Lyn is presented and discussed with inputs from various theoretical models.

The well resolved TESS data allow the identification of 23 frequencies in SZ Lyn. Apart from the fundamental radial mode of 8.296 d$^{-1}$ and its first two harmonics, we found 3 independent modes and 10 harmonics of the fundamental frequency. Though the ground based data are not up to the level of quality of TESS, most of the harmonics and one independent mode are present in the Mount Abu light curve also.  The presence of the dominant \textit{p}-mode of the radial fundamental at 8.296 d$^{-1}$, non-radial \textit{p}-modes at 14.535 d$^{-1}$, 32.620 d$^{-1}$ and a \textit{g}-mode at 4.584 d$^{-1}$ indicate that all possible oscillations can be simultaneously present in the same star. Additionally, the HELAS pulsation models give indication that 14.535 d$^{-1}$ and 32.620 d$^{-1}$ are close to spherical degree \textit{l} = 2 mode but this conclusion needs to be confirmed with more observational evidence.   

We attempted to refine the physical parameters of SZ Lyn through a rigorous analysis of stellar pulsation and evolutionary models.
Due to the large number of input parameters it is obvious that inconsistencies between different models make this approach difficult. 
Nevertheless, a suitable exploration of parameters can be a powerful means to determine approximate values for the effective temperature and mass of SZ Lyn. Our analysis points to a T$_{\rm eff} \approx$ 7500 K -- 7800 K and a mass in the range 1.7 M$_\odot$--2.0 M$_\odot$.

We also note that even with the availability of high-quality space-based data, UBVR ground-based observations are essential to constrain the different theoretical models.

\section*{Acknowledgements}

We acknowledge the anonymous referee for very pertinent suggestions.  We thank M. A. Dupret for providing non-adiabatic computation results. JA acknowledges the support of the Centre for International Cooperation in Science, 
Govt. of India for his work and stay at PRL when some of the work presented here was completed.  JA's earlier work at PRL was under a UN sponsored 
CSSTEAP (Centre for Space Science Technology and Education in Asia Pacific) program.  We thank the staff at the Mount Abu IR Observatory for their 
technical support during the course of this work. We are also very grateful to the staff of Arthur C Clarke Institute for their invaluable 
administrative support. We thank Prof U C Joshi and Prof H O Vats, for their support and useful discussion at various times through this work. SJ acknowledges the discussion with Dr. Dogus Ozuyar on evolutionary aspects. We are grateful to Dominic Bowman and Kosmas Gazeas for comments on the manuscript.

This work is supported by the Department of Space, Govt. of India. GH gratefully acknowledges funding through NCN grant 
2015/18/A/ST9/00578. 

"This paper makes use of data from the first public release of the WASP data (Butters et al. 2010) as provided by the WASP consortium and services at the NASA Exoplanet Archive, which is operated by the California Institute of Technology, under contract with the National Aeronautics and Space Administration under the Exoplanet Exploration Program."

This work has made use of data from the European Space Agency (ESA) mission {\it
Gaia} ({\tt https://www.cosmos.esa.int/gaia}), processed by the {\it
Gaia} Data Processing and Analysis Consortium (DPAC,
{\tt https://www.cosmos.esa.int/web/gaia/dpac/consortium}). Funding
for the DPAC has been provided by national institutions, in
particular the institutions participating in the {\it Gaia}
Multilateral Agreement.

This  paper  includes  data  collected  by  the  TESS  mission.  Funding  for  the  TESS  mission  is  provided by  the  NASA  Explorer  Program.  Funding  for  the  TESS Asteroseismic  Science  Operations Centre is provided by the Danish National Research Foundation (Grant agreement no.:  DNRF106),  ESA  PRODEX  (PEA  4000119301)  and Stellar  Astrophysics  Centre  (SAC) at Aarhus University. We thank the TESS team and staff and TASC/TASOC for their support of the present work.

JG's research has been partly supported by  the Spanish project \textit{PID 2019-109363GB-100}\\
\section*{Data Availability}

The TESS and WASP observations of SZ Lyn are publicly available from the respective  data archives.   The model data received from Dr Dupret and the other results computed by the authors are already included in the article. Intermediate results and the underlying observational data, from Mount Abu and APT, will be shared on reasonable request to the corresponding author.


\bibliographystyle{mnras}
\bibliography{reference} 











\bsp	
\label{lastpage}
\end{document}